\documentclass[%
nofootinbib,
nobibnotes,
 amsmath,amssymb, 
 reprint,
 prb,
]{revtex4-1}
\usepackage{graphicx}
\usepackage{dcolumn}
\usepackage{bm}

\begin{document}

\title{On the impact of a concave nosed axisymmetric body on a free surface}
\author{Varghese Mathai}
\thanks{Present address: \textit{Physics of Fluids Group, Faculty of Science and Technology, University of Twente, The Netherlands}}
\affiliation{Department of Mechanical Engineering, Indian Institute of Science, Bangalore 560012, India.}
\author{Raghuraman N. Govardhan}\thanks{Email for correspondence: raghu@mecheng.iisc.ernet.in}
\affiliation{Department of Mechanical Engineering, Indian Institute of Science, Bangalore 560012, India.}
\author{Vijay H. Arakeri}
\affiliation{Department of Mechanical Engineering, Indian Institute of Science, Bangalore 560012, India.}
\date{\today}

\begin{abstract}
We report on an experimental study of the vertical impact of a concave nosed axisymmetric body on a free surface. Previous studies have shown that bodies with a convex nose, like a sphere, produce a well defined splash with a relatively large cavity behind the model. In contrast, we find that with a concave nose, there is hardly a splash and the cavity extent is greatly reduced. This may be explained by the fact that in the concave nosed case, the initial impact is between a confined air pocket and the free surface unlike in the convex nosed case. From measurements of the unsteady pressure in the concave nose portion, we show that in this case, the maximum pressures are significantly lower than the classically expected ``water hammer'' pressures and also lower than those generally measured on other geometries. Thus, the presence of an air pocket in the case of a concave nosed body adds an interesting dimension to the classical problem of impact of solid bodies on to a free surface. 
\end{abstract}
\maketitle

The classical problem of rigid body- free surface interaction has many features, which make it interesting to study. Impact of solids in to water comprises a complicated series of events that occur both above and below the water surface, and depends on the configuration of the object used.  
There have been a large number of studies on this topic including some classical ones.\cite{Worthington1897,Richardson1948,Gilbarg1948} The primary motivation for these studies have been their relevance in applications like the landing of sea planes, spacecrafts and rocket parts on ocean surface, the slamming of a high speed ship with the water surface, supercavitation around rotating blades of hydraulic machinery and ship propellers operating near a free surface, and the flow field associated with air dropped underwater systems.

When an object, say a sphere, impacts vertically into water from air, the hydrodynamic phenomena of interest are the splash, cavities, and jets.\cite{Gilbarg1948} The first two form at early times after impact; whereas, jets appear at much later times. Even though the splash forms above the water surface, its dynamics can have an important influence on the extent of cavity formed in the wake of the object that in turn influences its underwater motion. A systematic and comprehensive investigation~\cite{May1952} was undertaken to study the influence of various parameters on splash and cavity formation accompanying vertical impact of solid objects on a water surface. The parameters considered in this study included the density and pressure of the atmosphere above water, the velocity, size and, to a limited extent, the nose shape of the objects. More recent studies on spheres and axisymmetric slender bodies have shown that some other parameters like surface properties and spin imparted to the object can have important influence on splash and cavity formation.\cite{Duez2007,Aristoff2009,Truscott2009a,Truscott2009b,Bodily2014}

\begin{figure} [!htbp]
\centering
\vspace{-0.2cm}
\includegraphics[width=0.27\textwidth]{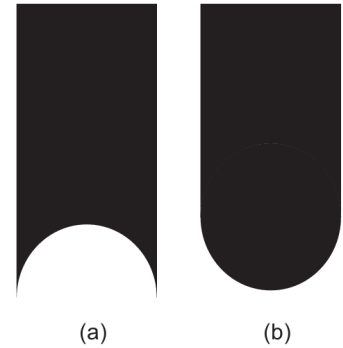}
\vspace{-0.3cm}
\caption{\footnotesize{Schematic of (a) concave and (b) convex nosed axisymmetric body.}}
\label{fig:1}
\vspace{-0.5cm}
\end{figure}

\begin{figure*} [!htbp]
\centering
\includegraphics[width=.8\textwidth]{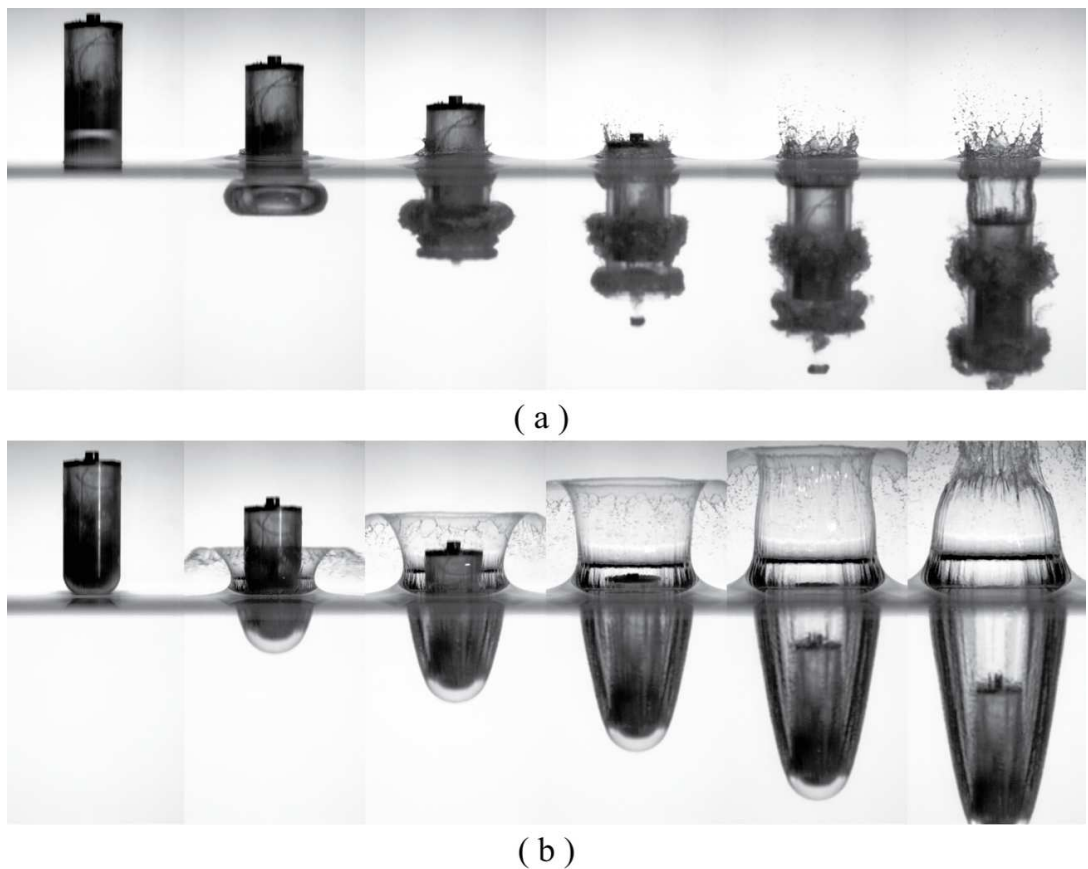}
\vspace{-0.3cm}\\
\caption{\footnotesize{Photographs comparing water impact of  (a) concave nose and (b) convex nose models at impact velocity of 9 m/s. Time between successive images is 2.4 ms. Both models had a diameter of 1 inch (25.4 mm). A movie showing the water impact of the two models is also included as supplementary material.\cite{Supplemental1}}} 
\label{fig:2}
\vspace{-0.3cm}
\end{figure*}

Other than the hydrodynamic features like splash and cavity formation, there is an important structural perspective to the water impact problem. The central question is what is the pressure experienced by the object at impact; the theoretical value is given by what is known as the ``water hammer'' pressure expressed as $\rho_{w} V_{0} c_{w}$, where $\rho_{w}$ is the water density, $V_{0}$ is the impact velocity, and $c_w$ is the speed of sound in water.\cite{Cook1928} As is evident from the expression, the water hammer pressure can take on very high values even with moderate impact velocities ($V_{0}$) of 10 m/s. However, a comprehensive experimental study\cite{Chuang1966} has shown that the maximum impact pressures are generally much lower than the water hammer pressure. The first theoretical explanation for the measured lower impact pressure was provided by taking into account the influence of the compressible air layer trapped between the falling object and the water surface.\cite{Verhagen1967} In the case of an impacting convex body, the existence of a thin trapped air layer has been experimentally demonstrated in a recent study,\cite{Hicks2012} where comparisons have also been made with a theoretical model proposed earlier.\cite{Hicks2010} 
Recent measurements\cite{VanNuffel2013} of impact pressures on a convex body in the form of a two-dimensional cylinder entering water at different velocities are of particular relevance to our present work.

As indicated previously, the influence of many parameters on the hydrodynamics following impact of a solid object on water have been investigated.
However, effect of one parameter, namely, the geometry of the solid object has received relatively little attention with most of the studies being on a sphere as a basic form of solid. In our present study, we have chosen a completely different form by investigating vertical impact of a concave nosed axisymmetric body (Fig.~\ref{fig:1}(a)) on a free surface. For this nose shape, we do high speed visualizations of the hydrodynamic features like splash and cavity formation, and also measure the impact pressure. We find that with this geometry, there are significant differences on all the above aspects in comparison to the convex nosed (Fig.~\ref{fig:1}(b)) and other bodies previously reported in the literature. 

The basic experimental setup consisted of a water tank and an electromagnetic model dropping arrangement that could be placed at arbitrary heights above the water surface to change the impact speed. The water tank used had dimensions of 70 cm x 70 cm x 140 cm and was filled with filtered water to a height of 80 cm for the tests. The maximum effective drop height using the present setup was 4.8 m, with corresponding maximum impact speed of 9.7 m/s. A schematic illustration of the concave nosed axisymmetric body studied is shown in Fig.~\ref{fig:1}(a), along with the reference convex nosed body in Fig.~\ref{fig:1}(b). In both cases, the models used were made up of a cylindrical after-body that followed the different nose shapes. The models used had a diameter of 1 and 2 inches (25.4 and 50.8 mm) and a length of 6 diameters. Impact velocities of up to about 10 m/s were studied. The models were released into the water tank with the help of an electromagnetic dropper that held on to a pointed metallic projection on the top end of the cylindrical Perspex models. The alignment of the models to ensure a vertical drop was done using a laser. Further, to ensure a stable vertical descent of the model, it was important that the model's centre of gravity be as low as possible. Hence, they were made hollow and lead weights inserted just aft of the nose to keep them bottom heavy.  In all cases reported here, the weight of the axisymmmetric models used was the same as a steel sphere of the same diameter. High speed flow visualizations of the model impact on water were done at 5000 frames per second using a Photron FASTCAM-SA5 camera with illumination from six 500-Watt Halogen lamps. Unsteady impact pressure measurements were done inside the 2 inch (50.8 mm) diameter concave nosed body using a PCB Piezotronics 112A21 pressure transducer that was mounted flush at the central part of the nose along the axis of the axisymmetric model. The transducer sensing diameter was 5.54 mm and has a usable frequency range of up to 250 kHz. Accuracy of peak impact pressure measurements was ensured by following recommended guidelines\cite{VanNuffel2013} for such measurements. 
\begin{figure} [!htbp]
\centering
\vspace{-0.2cm}
\includegraphics[width=0.45\textwidth]{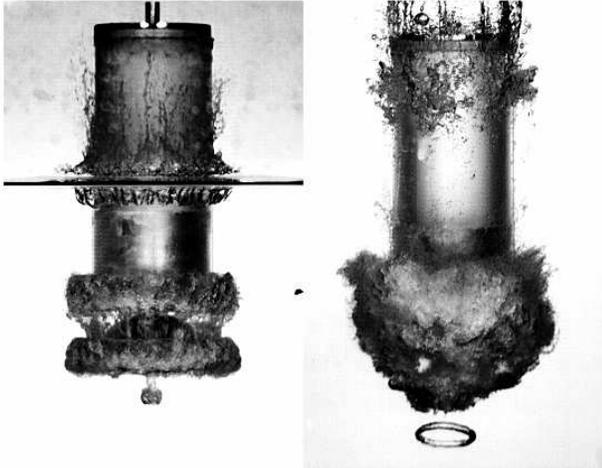}
\vspace{-0.3cm}
\caption{\footnotesize{Close-up photographs of a 2 inch (50.8 mm)  diameter concave nosed body after water impact at 7 m/s. One can see the ejection of a two-phase jet from the nose in (a), which eventually leads to the formation of a clear vortex ring ahead of the model at later times as seen in (b). Time after impact in (a) is 12 msec, while it is 37 msec in (b). A movie showing this visualization along with the corresponding pressure time trace, shown in Fig.~\ref{fig:4}, is included as supplementary material \cite{Supplemental2}.}}
\label{fig:3}
\vspace{-0.5cm}
\end{figure}

The time sequence of visualization showing the hydrodynamic aspects of impact for the concave and convex nose shapes is shown in Fig.~\ref{fig:2}. As may be seen from the visualizations, there are striking differences between the two nose shapes. In the well studied convex nose shape (Fig.~\ref{fig:2}(b)), from the instant of contact with the free surface, a thin sheet of water travels up the surface of the body and detaches itself from the solid surface. After detachment, this sheet termed as ``splash'' rises symmetrically keeping clear of the after-body and eventually forms a domed closed surface that prevents further passage of air into the air cavity. The concave nose shape is very different in its behavior, as is apparent from the sequence of images presented in Fig.~\ref{fig:2}(a). There is no perceptible splash formation upon impact, and there is no clear cavity originating from the nose section. There is also the appearance of a two-phase jet that pushes fluid out of the nose region as seen from the third image of the sequence. This jet leads to the formation of a clearly visible vortex ring seen in subsequent images. A close-up of the jet ejected from the nose cavity is shown in Fig.~\ref{fig:3}(a) along with a close-up of the vortex ring at later times in Fig.~\ref{fig:3}(b); both visualized with a larger 2'' (50.8 mm) model used for the pressure measurements.
We found that the differences observed in the impact characteristics of the two different nosed bodies leave a mark on the form and extent of cavity formed at later times. In the case of the convex nosed body, the cavity remains clear and the maximum length achieved is nearly twenty diameters; whereas, with the concave nosed body, the cavity is in the form of patchy two-phase regions with maximum length being only around two diameters. 
\begin{figure}
\centering
\vspace{-0.2cm}
\includegraphics[width=0.5\textwidth]{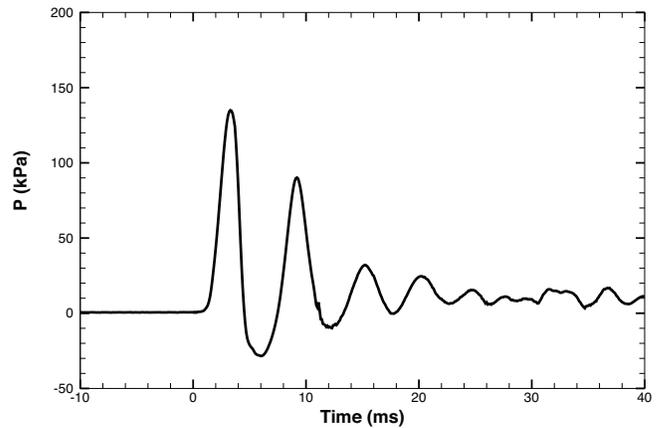}
\vspace{-0.3cm}
\caption{\footnotesize{Pressure time trace at the center of the concave nose along the axis of the axisymmetric model. The impact speed in this case was 7 m/s, and the model diameter was 2 inches (50.8 mm).}}
\label{fig:4}
\vspace{-0.5cm}
\end{figure}
\begin{figure}[h]
\centering
\includegraphics[width=0.5\textwidth]{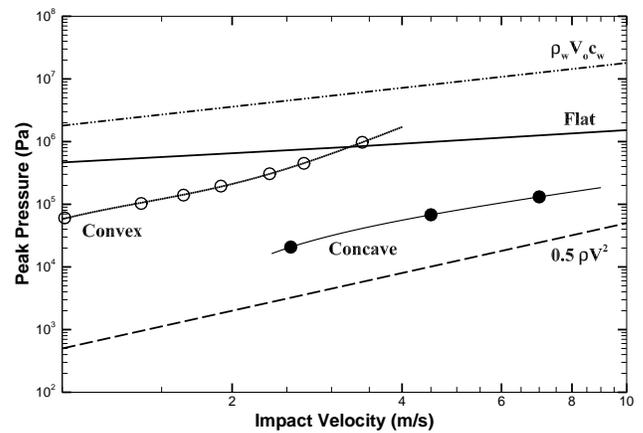}
\caption{\footnotesize{ Variation of the peak impact pressure for the present concave nosed body at different impact speeds. Also shown for comparison is the ``water hammer'' pressure ($\rho_{w} V_{0} c_{w} $), and the peak pressures at water impact for a convex cylindrical model\cite{VanNuffel2013} and a flat bottom model.\cite{Chuang1966} The peak pressures for the present concave nose model are two orders of magnitude lower than the water hammer pressures and about an order of magnitude lower than the convex and flat bottom model peak pressures. $\bullet$, Present concave nose; $\circ$, Convex (cylinder).\cite{VanNuffel2013}}} 
\label{fig:5}
\vspace{-0.5cm}
\end{figure}

We next present our results on the pressure measurements on the concave nosed body and compare them with other relevant values including those which are representative for a convex nosed body.\cite{VanNuffel2013} A typical time signal of measured pressure on the central part of the concave nose at impact speed of 7 m/s is shown in Fig.~\ref{fig:4}. The signal indicates that the nose experiences a sudden rise in pressure in the first few milliseconds following impact with peak pressure of about 140 kPa. This is followed by oscillations that damp out gradually in time. This type of pressure measurements have been done at a few different impact speeds for the concave nosed body and the peak pressures from each are plotted versus the impact speed in Fig.~\ref{fig:5}. The figure also shows the ``water hammer'' pressure and maximum pressures from literature for a flat bottom body\cite{Chuang1966} and a convex body,\cite{VanNuffel2013} the latter actually being for a cylinder falling with its axis parallel to the free surface. As is clear from the plot, the peak impact pressures for the present concave nose shape is about 2 orders of magnitude lower than the ``water hammer'' pressure, and more than an order of magnitude lower than the convex and the flat bottom model peak pressures at similar impact speeds. 

There are some aspects of the pressure measurements shown in figure 4, which merit further discussion. The observed oscillatory behavior is quite intriguing and it would be interesting to further study this behavior by investigating its dependence on some parameters like body size and particular shape of the nose concavity (parabolic, conical, flat etc.). It is also somewhat puzzling to note that even though the concave nosed body experiences significantly reduced impact pressures than the convex nosed one, the video (see supplementary material) shows that there is greater deceleration over time for the concave nosed body. However, this apparent discrepancy may not necessarily be contradictory once the timing characteristics of the initial pressure pulse are taken into account. We find from figure 4 that the pulse width at half maximum for the concave nosed body is about 2 milliseconds, whereas, it is about 0.045 milliseconds for the convex nosed body \cite{VanNuffel2013}. Therefore, on the average, the pressure of the concave nose, after the initial strike, remains at a higher level over a longer period than that observed for the convex nose. This is the most likely reason for the observed slower travel of the concave nosed body as compared to the convex one. 

In summary, we can state that there are some interesting dimensions to the problem of impact of solid bodies on a free surface with the concave nosed body studied here. Even at moderate impact velocities of the order of 10 m/s, the concave nosed body enters the free surface without forming a discernable splash. In addition, the length of the nose cavity formed and measured peak impact pressures are found to be smaller by an order of magnitude as compared to a convex nosed body. We also find that in the concave nosed body case there is a clear ejection of a two-phase jet from the fore body region that leads to the formation of a vortex ring ahead of the model. It would be of interest to study the effect of these changes on quantities like forces experienced by the model after impact and its subsequent underwater motion. The significant reductions in the cavity length and peak impact pressures could prove to be beneficial in some applications like landing of sea planes on ocean surface, dip coating technologies and development of air dropped underwater systems.

The authors thank Prof. G. Jagadeesh for use of the impact pressure transducer, and Narsing K. Jha and Nithiyaraj Munuswamy for help with preparation of the manuscript. 
\\
Copyright, American Institute of Physics, 2015. This article may be downloaded for personal use only. Any other use requires prior permission of the author and the American Institute of Physics. The following article has been modified from the version that appeared in Applied~Physics~Letters~\cite{Mathai2015impact}. The published version can be found at http://aip.scitation.org/doi/full/10.1063/1.4907555.

\end{document}